\documentclass[technote,10pt]{IEEEtran}
\usepackage{cite}
\usepackage{graphicx}
\usepackage{picinpar}
\usepackage{amsmath,amsfonts,amssymb}
\usepackage{amsthm}                 
\newtheorem{condition}{Condition}  
\usepackage{subfigure}
\usepackage{algorithm}
\usepackage{stfloats}
\usepackage{bm}
\usepackage{xcolor}
\usepackage{booktabs}
\usepackage{makecell}
\usepackage{mathrsfs}
\usepackage{epstopdf}
\DeclareMathAlphabet{\mathbbold}{U}{bbold}{m}{n}
\usepackage{xcolor}
\usepackage[colorlinks,bookmarksopen,bookmarksnumbered,citecolor=blue, linkcolor=blue, urlcolor=blue]{hyperref}
\usepackage{algpseudocode}
\usepackage{tikz}
\usetikzlibrary{spy,backgrounds}
\usepackage{flushend}

\newcommand{\Vcal}{\mathcal{V}}
\newcommand{\Ccal}{\mathcal{C}}

\begin{document}
	
	\title{Movable Antenna for Integrating Near-field Channel Estimation and Localization}
	\author{
		\IEEEauthorblockN{
			Chongjia Sun,
			Ziwei Wan, 
			Lipeng Zhu, 
			Zhenyu Xiao,
			Zhen Gao, and
			Rui Zhang,~\IEEEmembership{Fellow,~IEEE}}
		\vspace*{-3.0mm}

		\thanks{Chongjia Sun and Ziwei Wan are with the Yangtze Delta Region Academy, BIT (Jiaxing), Jiaxing 314019, China, and also with the School of Information and Electronics	and the Advanced Research Institute of Multidisciplinary Sciences, Beijing Institute of Technology, Beijing 100081, China(e-mail: \{3220242920, ziweiwan\}@bit.edu.cn).}
		\thanks{Z. Xiao is with the School of Electronic and Information Engineering,
			Beihang University, Beijing, China 100191 (e-mail: xiaozy@buaa.edu.cn).}
		\thanks{Zhen Gao is with the State Key Laboratory of CNS/ATM and the MIIT Key 	Laboratory of Complex-Field Intelligent Sensing, Beijing 100081, China, also	with Beijing Institute of Technology (BIT), BIT, Zhuhai 519088, China, also		with the Advanced Technology Research Institute, BIT (Jinan), Jinan 250307,		China, and also with the Yangtze Delta Region Academy, BIT (Jiaxing), Jiaxing 314019, China (e-mail: gaozhen16@bit.edu.cn).}
		\thanks{L. Zhu and R. Zhang are with the Department of Electrical and Computer Engineering, National University of Singapore, Singapore 117583 (e-mail: \{zhulp, elezhang\}@nus.edu.sg).}
		\thanks{The codes about this work are available at \url{https://github.com/ZiweiWan/Code-4-MA-Integrating-Near-field-CE-and-Localization.git}}
	}

	\maketitle
	
	\begin{abstract}
		Movable antenna (MA) introduces a new degree of freedom for future wireless communication systems by enabling the adaptive adjustment of antenna positions. Its large-range movement renders wireless channels transmission into the near-field region, which brings new performance enhancement for integrated sensing and communication (ISAC).
		This paper proposes a novel multi-stage design framework for broadband near-field ISAC assisted by MA. The framework first divides the MA movement area into multiple subregions, and employs the Newtonized orthogonal matching pursuit algorithm (NOMP) to achieve high-precision angle estimation in each subregion. Subsequently, a method called near-field localization via subregion ray clustering (LSRC) is proposed for identifying the positions of scatterers. This method finds the coordinates of each scatterer by jointly processing the angle estimates across all subregions. 
		Finally, according to the estimated locations of the scatterers, the near-field channel estimation (CE) is refined for improving communication performance.
		Simulation results demonstrate that the proposed scheme can significantly enhance MA sensing accuracy and CE, providing an efficient solution for MA-aided near-field ISAC. 
	\end{abstract}


	\section{Introduction}
	
	Massive or ultra-massive multiple-input multiple-output (MIMO) system is deemed to be a prerequisite for supporting high data rate and increasing number of served users in future wireless systems, especially for the envisioned 6G massive communication scenarios \cite{malong}. The increase in antenna aperture at the base station (BS) leads to a significantly longer Rayleigh distance, which broadens the near-field area of electromagnetic transmission \cite{xyz,TVT1}. In the context of near field, MIMO channels not only depend on the angles of scatterers but also closely coupled with their distance information, which creates many promising opportunities for communications and integrated sensing and communication (ISAC).
	
	However, massive or ultra-massive MIMO faces some disadvantages, especially in terms of hardware cost and signal processing complexity.
	The deployment of numerous radio frequency (RF) chains at the BSs induces unaffordable energy consumption. 
	Moreover, packing a large number of antennas into a compact space leads to heat dissipation issues \cite{zhu_model,framework,tuto}. 
	To address the challenges mentioned above, movable antenna (MA) has emerged as a promising paradigm shift of MIMO techniques. It opens up a new degree of freedom for enhancing wireless communication performance by enabling antennas to physically move at the transmitter or receiver \cite{lyn}. 
	Recently, this flexibility has been further extended to six-dimensional MAs, which jointly optimize 3D positions and 3D rotations to adapt to user spatial distributions \cite{shao_twc}.
	In this way, MA can use fewer antennas and RF chains to achieve the same or even better performance than traditional fixed antenna systems \cite{zhu_model,framework,tuto}.
	
	For harnessing the benefits of MA, obtaining the channel state information (CSI) for all candidate positions of the MA is essential.
	Prior research employed the compressive sensing (CS) algorithm for channel estimation (CE) in both fixed antenna and MA systems.
	For instance, a CS-based ISAC processing framework was proposed in \cite{wzw_isac} to reduce pilot overhead and recover high-dimensional channel information in mmwave massive MIMO.
	For MA systems, the authors in \cite{ma} proposed sequential estimation frameworks that successively estimate the angle of departure, angle of arrival, and path gains. In \cite{framework}, a joint estimation framework was proposed, for simultaneously	recovering all channel path parameters. 
	Besides CS-based approaches, tensor decomposition has also been exploited to leverage the structure of MA communication systems for channel parameter estimation \cite{tensor}.
	These methods have effectively reduced the CE overhead for far-field MA systems.
	On the other hand, research begins to shift towards MA-enabled near-field communications. In \cite{zhu_model}, the authors proposed algorithms for the joint optimization of antenna positions and beamforming in MA-aided communication systems. 
	A MA-aided near-field ISAC system was investigated in \cite{ding} to optimize the weighted sum performance of sensing and communication, while the MA-enabled near-field MU-MIMO transmission was studied in \cite{Pi} to enhance spatial multiplexing gains.
	However, the implementation of these optimization algorithms relies on the premise that the BS has already obtained accurate CSI, yet how to acquire this near-field CSI associated with MA remains a challenge.

	To address the challenges above, this paper proposes a multi-stage near-field MA ISAC framework.
	The main contributions of this paper are summarized as follows:
	\begin{itemize}
		\item We propose a subregion-based angle estimation scheme for near-field MA channel. 
		The MA movement area is divided into multiple subregions and the Newtonized orthogonal matching pursuit (NOMP) algorithm is introduced for angle estimation of each subregion, where the Newton's method is utilized to obtain high-precision angle estimates in the continuous angular domain.
		\item 
		We introduce a localization method, termed localization via subregion ray clustering (LSRC), to determine the coordinates of scatterers. 
		By jointing employing the angle estimates previously obtained, LSRC minimizes the sum of distances to the estimated rays from all subregions, thereby  determining the locations of scatterers for the sensing task. 
		\item 
		Utilizing the estimated scatterer locations from sensing, we solve the delays and complex gains to reconstruct the near-field channel for communication tasks.
		This method can improve CE performance compared to the initial CS-based CE, which validates the effectiveness of MA-aided ISAC.
	\end{itemize}
	
	\textit{Notation}: For a matrix \(\mathbf{A} \in \mathbb{C}^{m \times n}\), its $i$-th row and $j$-th column element is \(\left[\mathbf{A}\right]_{i,j}\). 
	The Frobenius norm of $\mathbf{A}$ is \(\|\mathbf{A}\|_{\text{F}}\). 
	For a column vector \(\mathbf{a} \in \mathbb{C}^m \) , its $i$-th element  is \(\left[\mathbf{a}\right]_{i}\).
	The $l_p$-norm of $\bf a$ is ${\left\| {\bf{a}} \right\|_p}$. The transpose, Hermitian, and inverse operations are denoted by $(\cdot)^{\text{T}}$, $(\cdot)^{\text{H}}$, and $(\cdot)^{-1}$, respectively.
	$\mathbf{I}_N$ denotes the $N$-order identity matrix.
	$\text{diag} \{\cdot\}$ denotes the diagonalization operation.
	$|\cdot|_\text{c}$ denotes the cardinality of a set.

	\begin{figure}[t]  
		\centering
		\includegraphics[width=0.5\textwidth]{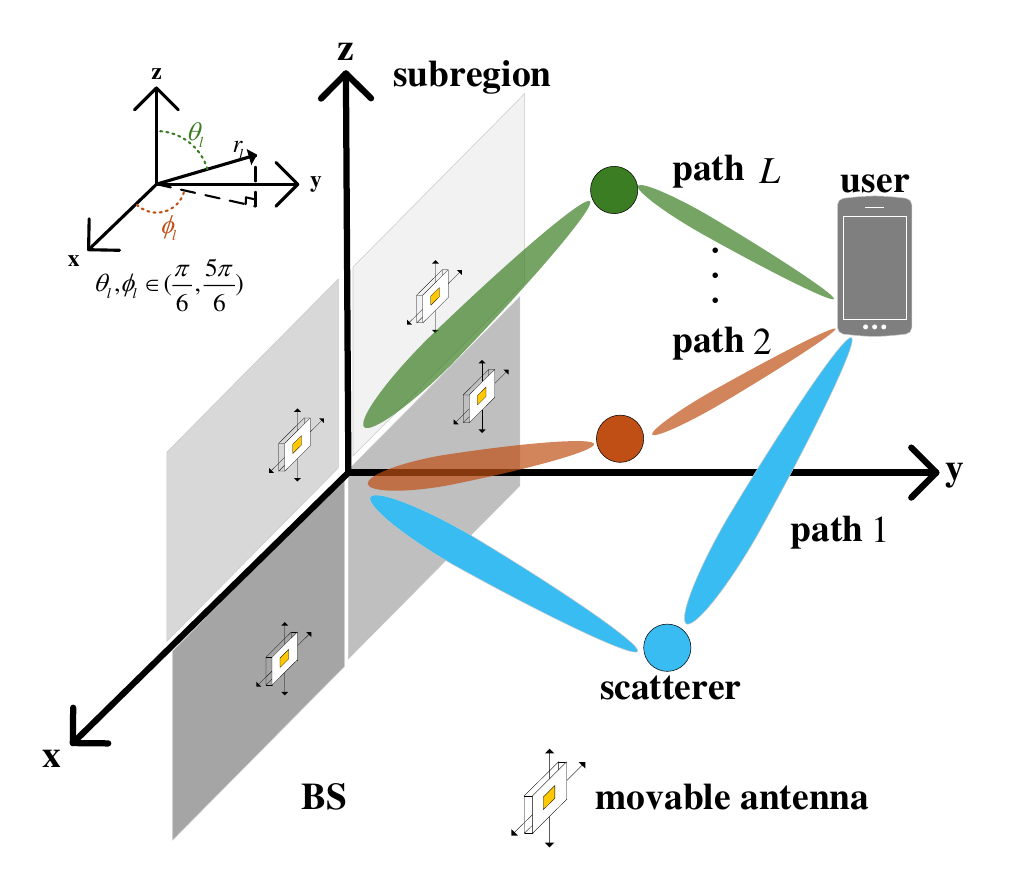}
		\caption{Near-field MA-aided communication system.}
		\label{fig1}
	\end{figure}

	\section{System Model}
	\subsection{System Description and Channel Model}
	
	As shown in Fig. \ref{fig1}, we consider an uplink near-field communication system comprising a BS equipped with $Q$ MAs and a single-antenna user. 
	It adopts orthogonal frequency division multiplexing (OFDM) with carrier frequency $f_{\rm c}$ and $K$ subcarriers $\{ f_1, f_2,..., f_K \}$.
	Each MA at the BS can move within the $x$-$o$-$z$ plane.
	Considering the mechanical constraints of stepper motors in practice, the movement of antennas is often limited to a finite set of discrete states \cite{shao_jsac}.
	Therefore, we consider a finite set of $N$ discrete  ports available for the MA \cite{framework}, whose coordinates are collected as
	$ \mathcal{R}= \{\mathbf{r}_1,\mathbf{r}_2,\dots,\mathbf{r}_{N}\}$ with ${\bf r}_n = [x_n,0,z_n]^{\rm T} \in \mathbb{R}^3$.
	The uplink near-field channel response is modeled as a superposition of $L$ resolvable paths induced by $L$ scatterers \cite{wzw_tcom}. That is
	\begin{equation}
		\mathbf{h}_k = \sum_{l=1}^{L} \beta_l \mathbf{a}(\mathbf{p}_l) e^{-j2\pi f_k \tau_l},
		\label{hk}
	\end{equation}
	where $\mathbf{h}_k \in \mathbb{C}^N$ is the channel vector at the $k$-th subcarrier, $\mathbf{p}_l = [r_l, \theta_l, \phi_l]^{\text{T}}$ is the spherical coordinate of the $l$-th scatterer (with $r_l, \theta_l, \phi_l$ being the radial distance, elevation angle, and azimuth angle, respectively), $\beta_l$ and $\tau_l$ represent complex gain and delay associated with the $l$-th scatterer, respectively, and $\mathbf{a}(\mathbf{p}_l) \in \mathbb{C}^{N}$ is the near-field steering vector.
	The $n$-th element of ${\bf a}({\bf p}_l)$ is given by
	$[\mathbf{a}(\mathbf{p}_l)]_n = e^{-j\frac{2\pi}{\lambda}(\Vert \mathbf{r}_n - \mathbf{s}_l \Vert_2 - r_l)}$,
	where $\lambda$ is the carrier wavelength, and $\mathbf{s}_l=[r_l \text{sin}\theta_l\text{cos}\phi_l,r_l \text{sin}\theta_l\text{sin}\phi_l,r_l \text{cos}\theta_l]^{\text{T}}  \in \mathbb{R}^3$ is the Cartesian coordinate corresponding to ${\bf p}_l$.

	\subsection{Signal Model}
	As shown in Fig. \ref{fig1}, we divide the entire planar region (i.e., discretized antenna moving region) into $Q$ non-overlapping subregions. Each subregion is equipped with a single MA.
	We assume  $N_\text{T}$ time slots for CE/sensing. In the $t$-th time slot ($t=1, \dots, N_\text{T}$), the MA in each subregion will move to a new port that has not been previously visited.
	Let the coordinates of visited ports of the $q$-th subregion be denoted as $\mathcal{I}_{q} \subset \{1,2,...,N\}$ 
	($q=1,2,\dots,Q$) with  $|\mathcal{I}_q|_\text{c}=N_\text{T}$ and $\mathcal{I}_q \cap \mathcal{I}_{q'} = \varnothing$ for $q \neq q'$.
	By assuming that the normalized pilot signal equal to one is transmitted, the received signal of the $q$-th subregion at the $k$-th subcarrier $\mathbf{y}^{(q)}_k \in \mathbb{C}^{N_\text{T}}$ is defined as
	\begin{equation}
		\mathbf{y}_k^{(q)} = \mathbf{S}_q^{\text{pos}} \mathbf{h}_k  + \mathbf{z}_k^{(q)},
	\end{equation}
	where $\mathbf{S}_q^{\text{pos}}= \left[ {\bf I}_N \right]_{{\cal I}_q,:} \in \{0,1\}^{N_\text{T} \times N}$ denotes the port selection matrix, $\mathbf{z}_k^{(q)} \sim \mathcal{CN}(\mathbf{0}, \sigma^2 \mathbf{I}_{N_\text{T}}) $ is the additive white Gaussian noise (AWGN) vector with power $\sigma^2$.
	Furthermore, we assume that only $K_\text{c}$ out of $K$ subcarriers are used for pilot transmission.
	The set of used pilot subcarriers is denoted as $\mathcal{J} \subset \{1,2,\dots,K\}$ with $|\mathcal{J}|_\text{c}=K_\text{c}$. 
	Thus, the pilot matrix received by the $q$-th subregion $\mathbf{Y}^{(q)} \in \mathbb{C}^{N_\text{T} \times K_\text{c}}$ is given by
	\begin{equation}
		\begin{aligned}
			\mathbf{Y}^{(q)}
			&=\mathbf{S}_q^{\text{pos}} \mathbf{H}  \mathbf{S}^{\text{sc}}+
			\mathbf{Z}^{(q)},
		\end{aligned}
		\label{Yq}
	\end{equation}
	where $\mathbf{S}^{\text{sc}}= \left[ {\bf I}_K \right]_{:,{\cal J}} \in \{0,1\}^{K\times K_\text{c}}$ is the subcarrier selection matrix,
	$\mathbf{H} = \left[\mathbf{h}_1,\mathbf{h}_2,\dots,\mathbf{h}_K \right] \in  \mathbb{C}^{N \times K}$,
	and $\mathbf{Z}^{(q)}=\left[\mathbf{z}_1^{(q)},\mathbf{z}_2^{(q)},\dots,\mathbf{z}_K^{(q)} \right] \mathbf{S}^{\text{sc}} \in  \mathbb{C}^{N_\text{T} \times K_\text{c}}$.
	According to \eqref{hk}, $\mathbf{H}$ can be rewritten as
	\begin{equation}
		\mathbf{H} = \mathbf{A} \mathbf{B} \mathbf{F}^{\text{T}},
		\label{H}
	\end{equation}
	where
	$\mathbf{A}=\left[\mathbf{a}(\mathbf{p}_1),\mathbf{a}(\mathbf{p}_2),\dots,\mathbf{a}(\mathbf{p}_L) \right] \in \mathbb{C}^{N \times L}$,
	$ \mathbf{B} = \text{diag} \left\{ \beta_1, \beta_2, \dots, \beta_L \right\} $,
	$\mathbf{F}=\left[\mathbf{f}(\tau_1),\mathbf{f}(\tau_2),\dots,\mathbf{f}(\tau_L) \right] \in \mathbb{C}^{K \times L}$, and $\mathbf{f}(\tau_l) = \left[e^{-j2\pi f_1 \tau_l},e^{-j2\pi f_2 \tau_l},...,e^{-j2\pi f_K \tau_l}\right]^{\text{T}} \in \mathbb{C}^K$.

	\section{Proposed MA-based Near-field ISAC Scheme}
	This section demonstrates the proposed near-field ISAC scheme with MAs, focusing on the localization of scatterers and the sensing-aided CE.	
	\subsection{Sparse Representation and Problem Formulation}	
	To utilize the distance-dependent phase shifts in near-field CE, existing studies  adopt polar-domain representation of near-field channels by sampling the angles and radial distance at the same time.
	However, it is reported that for a certain pair of angles, the near-field steering vectors corresponding to different radial distances are highly correlated \cite{xing}, making them indistinguishable during CE. 
	On the other hand, the region division suppresses the near-field effect within each subregion. 
	Dividing the region reduces the Rayleigh distance quadratically (i.e., to $1/4$ for a $Q=2\times2$ division). This effectively validates the plane wave assumption within each subregion for the considered scatterer range.
	These motivate us to apply the dictionary in only the angular domain to represent the channel.
	Specifically, we uniformly discretize the elevation angle and azimuth angle into $G_\theta$ and $G_\phi$ samples, respectively \cite{TVTWZW}. That is
	\begin{align}
		\bar{\theta}_{g_1}&=\frac{\pi}{6}+\frac{2\pi}{3}\frac{g_1}{G_\theta}\text{, }g_1 = 1,2,...G_\theta,\\
		\bar{\phi}_{g_2}&=\frac{\pi}{6}+\frac{2\pi}{3}\frac{g_2}{G_\phi}\text{, }g_2 = 1,2,...G_\phi.
	\end{align}
	Therefore, the total number of angular-domain samples is $G = G_\theta  G_\phi$, and each sample corresponds to a spatial location $\bar{\mathbf{p}}_g=\left[r_\text{fix}, \bar{\theta}_{g_1}, \bar{\phi}_{g_2} \right]$,
	where $g=g_1+(g_2-1)G_\theta \in [1,G]$ and  $r_\text{fix}$ is a predefined fixed radial distance.
	On this basis, we generate near-field steering vector  $\mathbf{a}(\bar{\mathbf{p}}_g)$ as the codeword and thus the dictionary matrix $\bar{\mathbf{A}}\in \mathbb{C}^{N \times G}$ can be obtained as $\bar{\mathbf{A}}=\left[\mathbf{a}(\bar{\mathbf{p}}_1),\mathbf{a}(\bar{\mathbf{p}}_2),\dots,\mathbf{a}(\bar{\mathbf{p}}_G) \right]$.
	We can use such a dictionary to represent the actual near-field channel as $\mathbf{H} \approx  \bar{\mathbf{A}}\bar{\mathbf{X}}$, where $\bar{\mathbf{X}}\in \mathbb{C}^{G \times K}$ is the sparse coefficient matrix. 
	Substituting this into \eqref{Yq} yields
	\begin{equation}
		\mathbf{Y}^{(q)} \approx
		\mathbf{S}_q^{\text{pos}} \bar{\mathbf{A}} \bar{\mathbf{X}} \mathbf{S}^{\text{sc}}+
		\mathbf{Z}^{(q)}.
		\label{cs}
	\end{equation}
	
	Note that under the condition of finite bandwidth in practice, different columns of $\bar{\mathbf{X}}$ share a common sparse pattern, which makes \eqref{cs}  a multiple-measurement vector (MMV) problem.
	Accordingly, we transform the CE problem into a simultaneous sparse signal recovery problem, i.e.,
	\begin{equation}
		\begin{aligned}
			& \quad \arg\min_{\bar{\mathbf{X}}}
			\| \mathbf{Y}^{(q)} -
			\mathbf{S}_q^{\text{pos}} \bar{\mathbf{A}} \bar{\mathbf{X}} \mathbf{S}^{\text{sc}}\|_{\text{F}}^2,\\
			\text{s.t.} & \quad \|\bar{\mathbf{X}}_{:,k}\|_0=L_{\text{pre}},\text{ }k=1,2,...,K,\\
			& \quad \text{supp}\{\bar{\mathbf{X}}_{:,1}\}=\text{supp}\{\bar{\mathbf{X}}_{:,2}\}=...=\text{supp}\{\bar{\mathbf{X}}_{:,K}\},
		\end{aligned}
		\label{SRP}
	\end{equation}
	where $L_{\text{pre}}$ is a predefined integer.
	On the other hand, the sensing task can be cast into solving the locations (coordinates) of scatterers $\{\mathbf{p}_l\}_{l=1}^L$ based on received signal \eqref{Yq}.
	
	The initial CE problem \eqref{SRP} can be solved by many existing algorithms \cite{ma}, so the details are omitted in this paper for brevity. Instead, we focus only on the sensing task to obtain the locations of scatterers.
	It is worth noting that the locations of scatterers obtained from the sensing task can also benefit CE, which will be discussed later in this paper.

	\begin{algorithm}[t]
		\color{black}
		\caption{Subregion Angle Estimation Based on NOMP}
		\label{alg:nomp_mmv}
		\begin{algorithmic}[1] 
			
			\Require
			Received signals $\{{\bf Y}^{(q)}\}_{q=1}^Q$,
			dictionary matrix $\bar{\mathbf{A}} $,
			predefined integer $L_{\text{pre}}$, 
			and the number of Newton refinement iterations $R$.
			
			\Ensure
			Estimated elevation and azimuth angles matrix $\hat{\mathbf{\Theta}}$, $\hat{\mathbf{\Phi}} \in \mathbb{R}^{L_{\text{pre}}\times Q}$.
			
			\For{$q=1,2,...,Q$}
			\State Initialization: $\mathbf{Y}_{\text{res}} \leftarrow \mathbf{Y}^{(q)}$,
			$\mathcal{P}_q \leftarrow \emptyset$,
			$\hat{\mathbf{A}} \leftarrow [\ ]$.
			
			\For{$\tilde{l}=1,2,...,L_{\text{pre}}$}
			\State $g^* = \text{arg} \mathop{\text{max}}_{g}\left\|\mathbf{Y}_\text{res}^{\text{H}}\left(\mathbf{S}_q^{\text{pos}} \mathbf{a}(\bar{\mathbf{p}}_g)\right)\right\|_2^2$.
			\label{line:init1}
			
			\State Refinement initialization: $\hat{\mathbf{p}}_{0} \leftarrow \bar{\mathbf{p}}_{g^*}$.
			\label{line:init2}
			\For{$r=1,2,...R$}\label{line:nt1}
			\State The $r$-th iteration for Newton's method \cite{nomp}:
			$\hat{\mathbf{p}}_{r}=\hat{\mathbf{p}}_{r-1} - \left( \left. \mathbf{\Xi}(\mathbf{p}) \right|_{\mathbf{p}=\hat{\mathbf{p}}_{r-1}} \right)^{-1} \left. \nabla {J}(\mathbf{p}) \right|_{\mathbf{p}=\hat{\mathbf{p}}_{r-1}}$,
			where $\nabla {J}$ and $\mathbf{\Xi}$ are given in \eqref{jh1} and \eqref{jh2}.\label{line:nt_ref}
			\EndFor
			\State Update:  $\hat{\mathbf{p}}=[r_\text{fix},\hat{\theta}_q^{(\tilde{l})}, \hat{\phi}_q^{(\tilde{l})}] \leftarrow \hat{\mathbf{p}}_{R}$.
			\State $[\hat{\mathbf{\Theta}}]_{\tilde{l},q} = \hat{\theta}_q^{(\tilde{l})}$, $[\hat{\mathbf{\Phi}}]_{\tilde{l},q} = \hat{\phi}_q^{(\tilde{l})}$.
			\State  $\hat{\mathbf{A}} \leftarrow [\hat{\mathbf{A}}, \mathbf{a}(\hat{\mathbf{p}})]$.
			\State 
			$\tilde{\mathbf{X}} = (\mathbf{\Gamma}^{\text{H}} \mathbf{\Gamma})^{-1} \mathbf{\Gamma}^{\text{H}} \mathbf{Y}^{(q)}$, 
			where $\mathbf{\Gamma} =\mathbf{S}_q^{\text{pos}} \hat{\mathbf{A}}$. 
			\State 
			$\mathbf{Y}_{\text{res}} = \mathbf{Y}^{(q)} - \mathbf{\Gamma} \tilde{\mathbf{X}}.$\label{line:nt2}
			\EndFor
			\EndFor
			
		\end{algorithmic}
	\end{algorithm}

	\subsection{Angle Estimation Based on NOMP}

	We utilize NOMP for off-grid angle estimation for the sensing task, which is summarized in Algorithm  \ref{alg:nomp_mmv}. Specifically, we first obtain the on-grid coarse angle estimates by correlation operation in MMV-OMP, as done in steps \ref{line:init1} and \ref{line:init2}.
	Then we add a Newtonized refinement stage to improve the angle estimation performance (steps \ref{line:nt1}--\ref{line:nt2}). To this end, an objective function ${J}(\mathbf{p})$ taking MMV into account is first defined as 
	\begin{equation}
		{J}(\mathbf{p})=
		||\left(\mathbf{S}_q^{\text{pos}} \mathbf{a}(\mathbf{p})\right)^{\text{H}} \mathbf{Y}_\text{res}||_\text{F}^2=||\mathbf{a}_q(\mathbf{p})^\text{H}||_\text{F}^2,
	\end{equation}
	where $\mathbf{a}_q(\mathbf{p})=\mathbf{Y}_\text{res}^{\text{H}}\mathbf{S}_q^{\text{pos}} \mathbf{a}(\mathbf{p})$, $\mathbf{Y}_\text{res}$ denotes the current residual matrix.
	The Newton's method is then applied $R$ times to refine the on-grid estimate as done in step \ref{line:nt_ref}, where
	$\mathbf{\Xi}(\mathbf{p})$ is the Hessian matrix of ${J}(\mathbf{p})$, and $\nabla {J}(\mathbf{p})$ is the gradient vector. Particularly, 
	\begin{align}
		\label{jh1}
		\left[\nabla {J}(\mathbf{p})\right]_i &= 2 \text{Re}\{\mathbf{a}_q(\mathbf{p})^{\text{H}} 
		\frac{\partial \mathbf{a}_q(\mathbf{p})}{\partial p_i}\}, \\
		\begin{split}
			\left[\mathbf{\Xi}(\mathbf{p})\right]_{i,j} &= 2 \text{Re}\Big\{ \Big(\frac{\partial \mathbf{a}_q(\mathbf{p})}{\partial p_i}\Big)^{\text{H}}  \frac{\partial \mathbf{a}_q(\mathbf{p})}{\partial p_j} + \mathbf{a}_q(\mathbf{p})^{\text{H}} \frac{\partial^2 \mathbf{a}_q(\mathbf{p})}{\partial p_i \partial p_j}\Big\},
		\end{split}
		\label{jh2}
	\end{align}
	where 
	$p_i, p_j \in \{ \theta, \phi \}$.
	By executing the NOMP algorithm on each subregion, we obtain, for each detected path, $Q$ angle estimates from $Q$ different subregions.
	Note that unlike the far-field scenarios where the angle of one scatterer is the same for all subregions, the near-field effect results in the diverse angles across different subregions. This diversity will help to localize scatterers, as detailed in the following.
	

	\subsection{Near-Field LSRC}
	Theoretically, multiple rays originating from different subregions and pointing towards the same scatterer should intersect at a single point in space which should be the scatterer's location.
	In practice, due to noise and estimation errors, the NOMP algorithm might detect fake paths in different subregions or different subregions may point to different scatterers in the same estimated path. To address these problems, we propose an SRC step for the obtained angle estimates before performing localization.
	We collect all the obtained angle estimates as
	$\mathcal{V}_{\tilde{l}} =\{\tilde{\mathbf{v}}_1^{(\tilde{l})},\tilde{\mathbf{v}}_2^{(\tilde{l})},\dots,\tilde{\mathbf{v}}_Q^{(\tilde{l})}\}$ where $\tilde{\mathbf{v}}_q^{(\tilde{l})}= [\text{sin}\hat{\theta}_q^{(\tilde{l})}\text{cos}\hat{\phi}_q^{(\tilde{l})}, \text{sin}\hat{\theta}_q^{(\tilde{l})}\text{sin}\hat{\phi}_q^{(\tilde{l})}, \text{cos}\hat{\theta}_q^{(\tilde{l})}] \in \mathbb{R}^3$
	is the direction vector (DV) of the estimated path, ${\tilde{l}} = 1,2,...,{L_{\text{pre}}}$.
	The objective of SRC is to divide each of $\{\mathcal{V}_{\tilde{l}}\}_{\tilde{l}=1}^{L_{\text{pre}}}$ into different clusters, each accounting for one channel scatterer.
	To this end, we first introduce the following {\bf Condition 1} for a DV set.
	
	\begin{algorithm}[t]
		\caption{Localization via Subregion Ray Clustering (LSRC)}
		\label{alg:clustering}
		\begin{algorithmic}[1] 
			
			\Require
			$\{\Vcal_{\tilde{l}}\}_{\tilde{l}=1}^{L_{\text{pre}}}$, {\bf Condition 1} ($\alpha_\text{th}$).
			
			\Ensure
			The total number of clusters $N_\text{clu}$,
			and the estimate scatterers' spherical coordinate $\{\tilde{\mathbf{p}}_\iota\}_{\iota=1}^{N_\text{clu}}$.
			
			\State \textbf{Initialization:} $\iota=1$.
			
			\For{$\tilde{l} = 1$ to $L_{\text{pre}}$}\label{clu1}
			\State $\Vcal_\text{temp} \leftarrow \Vcal_{\tilde{l}}$.
			
			\While{$\Vcal_\text{temp} \ne \emptyset$}
			\State Select an arbitrary $\mathbf{v} \in \Vcal_\text{temp}$, and
			$\Ccal_\text{new} \leftarrow  \{\mathbf{v}\}$, $\Vcal_\text{temp} \leftarrow \Vcal_\text{temp} \setminus \{\mathbf{v}\}$.
			
			\For{each $\mathbf{v'} \in \Vcal_\text{temp}$}
			\If{$\Ccal_\text{new} \cup \{\mathbf{v'}\}$ satisfies {\bf Condition 1}}
			\State $\Ccal_\text{new} \leftarrow \Ccal_\text{new} \cup \{\mathbf{v'}\}$, $\Vcal_\text{temp} \leftarrow \Vcal_\text{temp} \setminus \{\mathbf{v'}\}$.
			\EndIf
			\EndFor\label{clu2}

			\If{$|\Ccal_\text{new}|_\text{c} \geq 2$}\label{cond2}
			\State ${{\cal C}_\iota } = {{\cal C}_{{\text{new}}}}$, $C_\iota=|\Ccal_\iota|_\text{c}$, $\iota=\iota+1$.
			\EndIf
			\EndWhile

			\EndFor
			
			\State $N_\text{clu} \leftarrow \iota - 1$.
			
			\For{$\iota=1,2,...,N_\text{clu}$}\label{line:Lf1}
			\State For the $\iota$-th cluster $\mathcal{C}_\iota=\{\mathbf{v}_1,\mathbf{v}_2,...,\mathbf{v}_{C_\iota}\}$, obtain the LS solution via \eqref{loc}, and further obtain its corresponding spherical coordinate $\tilde{\mathbf{p}}_\iota$.
			\EndFor \label{line:Lf2}
		\end{algorithmic}
	\end{algorithm}
	
	
	\begin{condition} \label{cond:dv_angle}
		A DV set $\mathcal{S}$ satisfies 
		\begin{equation*}
			{\rm arccos}(\mathbf{v}_{i}^{\text{T}} \mathbf{v}_{j}) < \alpha_{{\rm th}}, \quad
			\forall \mathbf{v}_{i}, \mathbf{v}_{j} \in \mathcal{S},
		\end{equation*}
		where $\alpha_{\rm th}$ is a predefined positive threshold.
	\end{condition}
	
\begin{figure}[t]  
		\centering
		\includegraphics[width=0.5\textwidth]{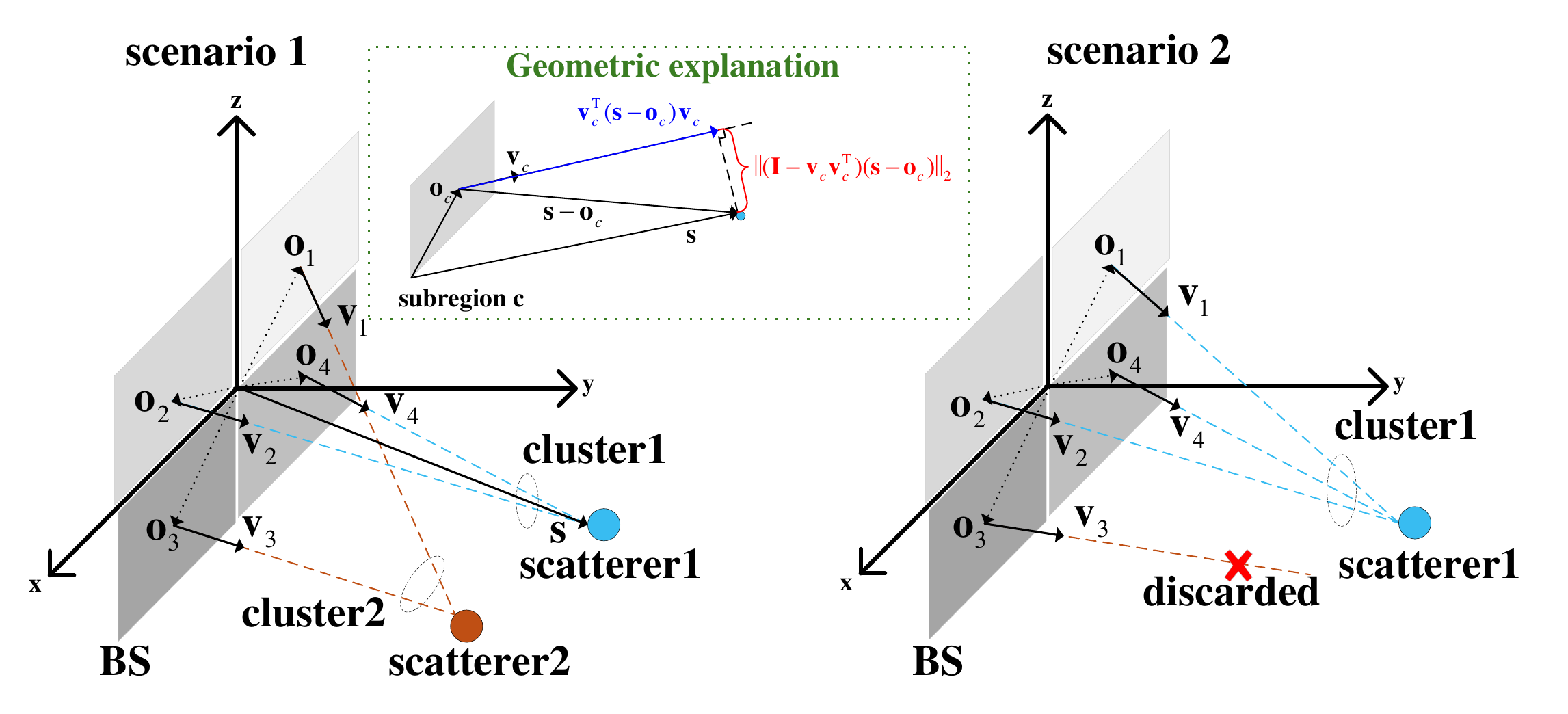}
		\caption{Near-field localization via subregion ray clustering.}
		\label{fig2}
	\end{figure}

	The clustering process is summarized in steps \ref{clu1}--\ref{clu2} of Algorithm \ref{alg:clustering}.
	In a nutshell, $Q$ DVs for each estimated path will be classified into several valid clusters. Each valid cluster contains similar DVs (controlled by $\alpha_\text{th}$) that corresponds to one certain channel scatterer, as shown in scenario 1 of Fig. \ref{fig2}, and will be used for subsequent localization. 
	On the other hand, DVs that fail to form clusters will be discarded, as shown in scenario 2 of Fig. \ref{fig2}.
	
	Without loss of generality, we consider one DV set from clustering results, $\mathcal{C}=\{\mathbf{v}_1,\mathbf{v}_2,...,\mathbf{v}_C\} $ ($C \leq Q$) to elaborate the subsequent localization method, as shown in steps \ref{line:Lf1}-\ref{line:Lf2} of Algorithm \ref{alg:clustering}.
	We construct a cost function $E_\text{loc}(\mathbf{s})$ as the sum of the squared distances between a point, denoted by the Cartesian coordinate $\bf s$, and all DVs in $\mathcal{C}$, which is given by
	\begin{equation}
		E_\text{loc}(\mathbf{s})=\sum_{c=1}^{C} \|(\mathbf{I}_3 - \mathbf{v}_c \mathbf{v}_c^{\text{T}})(\mathbf{s} - \mathbf{o}_c)\|_2^2,
	\end{equation}
	where $\mathbf{o}_c \in \mathbb{R}^3$ is the center of the subregion corresponding to the $c$-th DV in $\cal C$,
	and the geometric explanation of the distance representation is shown in Fig. \ref{fig2}.
	Our objective is to find the position that minimizes this cost function.  
	By utilizing the least square (LS) method, i.e., letting $\frac{d}{d\mathbf{s}}E_\text{loc}(\mathbf{s})=\mathbf{0}$, we have
	\begin{equation}
		\mathbf{\Gamma}
		{\mathbf{s}}=
		\boldsymbol{\gamma}
		\label{loc}
	\end{equation}
	where $\mathbf{\Gamma}=\sum_{c=1}^{C} (\mathbf{I}_3 - \mathbf{v}_c \mathbf{v}_c^{\text{T}})$ and $\boldsymbol{\gamma}=\sum_{c=1}^{C}(\mathbf{I}_3 - \mathbf{v}_c \mathbf{v}_c^{\text{T}}) \mathbf{o}_c$.
	Then we utilize LS to solve \eqref{loc}.
	Note that when $C=1$, $\mathbf{\Gamma}$ is not invertible\footnote{When $C=1$, the matrix simplifies to $\mathbf{\Gamma} = \mathbf{I}_3 - \mathbf{v}_1\mathbf{v}_1^\text{T}$. Utilizing the determinant identity $\det(\mathbf{I}_m - \mathbf{AB}) = \det(\mathbf{I}_n - \mathbf{BA})$ with $\mathbf{A} = \mathbf{v}_1$ and $\mathbf{B} = \mathbf{v}_1^\text{T}$, the determinant becomes $\det(\mathbf{\Gamma}) = \det(1 - \mathbf{v}_1^\text{T}\mathbf{v}_1)$. Since $\mathbf{v}_1$ is normalized, $\det(\mathbf{\Gamma}) = 0$, confirming singularity.}.
	Thus, we set the minimum cardinality of a valid cluster to $2$, as done in step \ref{cond2} of Algorithm \ref{alg:clustering}.
	
	\subsection{Sensing-assisted CE}
	
	Based on the sensing results obtained from Algorithm \ref{alg:clustering}, we focus on reconstructing the refined channels at all subcarriers.
	We combine all received signals to obtain $\mathbf{Y} = \left[\mathbf{Y}^{(1)};\mathbf{Y}^{(2)};\dots;\mathbf{Y}^{(Q)} \right] 
	\in  \mathbb{C}^{M \times K_\text{c}}$, where $M=Q N_\text{T}$. 
	Let the set of all visited ports of all MA be denoted by the union $\mathcal{I} = \mathcal{I}_{1} \cup \mathcal{I}_{2} \cup \dots \cup \mathcal{I}_{Q}$, and
	$\mathbf{S}^{\text{pos}}= \left[ {\bf I}_N \right]_{{\cal I},:} \in \mathbb{C}^{M \times N}$.
	$\mathbf{Y}$ can be written as
	\begin{equation}
		\mathbf{Y}=\mathbf{\Psi} \mathbf{X}_\text{sam} + \mathbf{Z},
		\label{Y}
	\end{equation}
	where $\mathbf{Z}=\left[ \mathbf{Z}^{(1)};\mathbf{Z}^{(2)};\dots;\mathbf{Z}^{(Q)} \right]
	\in \mathbb{C}^{M \times K_\text{c}}$, 
	$\mathbf{\Psi}=
	\mathbf{S}^{\text{pos}} \mathbf{A}
	\in \mathbb{C}^{M \times L}$,
	and $\mathbf{X}_\text{sam} = \mathbf{B} \mathbf{F}^{\text{T}} \mathbf{S}^\text{sc} \in \mathbb{C}^{L \times K_\text{c}}$.
	To estimate the delay components, we first replace the actual near-field steering matrix, $\mathbf{A}$, by the estimated version, $\hat{\mathbf{A}}$ from sensing results, where $\hat{\mathbf{A}}=[ 
	\mathbf{a}(\tilde{\mathbf{p}}_1), \mathbf{a}(\tilde{\mathbf{p}}_2),\dots,
	\mathbf{a}(\tilde{\mathbf{p}}_{N_\text{clu}})]$, and then obtain
	$\hat{\mathbf{X}}_\text{sam} = (\hat{\mathbf{\Psi}}^{\text{H}} \hat{\mathbf{\Psi}})^{-1} \hat{\mathbf{\Psi}}^{\text{H}} \mathbf{Y} \in \mathbb{C}^{N_\text{clu} \times K_\text{c}}$, where $\hat{\mathbf{\Psi}}=\mathbf{S}^{\text{pos}} \hat{\mathbf{A}}$.
	The $\iota$-th row of $\mathbf{\hat X}_\text{sam}$ can be modeled as
	\begin{equation}
		\hat{\mathbf{x}}_\text{sam}^{(\iota)} \approx
		\tilde{\beta}_{\iota}
		\mathbf{f}_\text{sam}(\tilde{\tau}_{\iota}),
		\label{xsam}
	\end{equation}
	where $\mathbf{f}_\text{sam}(\tilde{\tau}_{\iota})=\mathbf{f}(\tilde{\tau}_{\iota}) \mathbf{S}^{\text{sc}}  \in \mathbb{C}^{K_\text{c}}$, 
	$\tilde{\beta}_\iota$ and $\tilde{\tau}_\iota$ are the effective complex gain and delay for the $\iota$-th estimated scatterer, respectively.
	We uniformly discretize possible delay range into $G_\tau$ samples, ${\cal T} = \left\{ {{{\bar \tau }_1},{{\bar \tau }_2},...,{{\bar \tau }_{{G_\tau }}}} \right\}$, where ${{\bar \tau }_d} = d\tau_\text{max}/{G_\tau }$ ($d = 1,2,...,G_{\tau}$), and $\tau_\text{max}$ is the maximum delay spread.
	For each delay $\bar{\tau}_d$ in $\mathcal{T}$, we can obtain
	\begin{equation}
		\bar{\beta}_d^{(\iota)}=
		\frac{(\hat{\mathbf{x}}_\text{sam}^{(\iota)})^{\text{H}} \mathbf{f}_\text{sam}(\bar{\tau}_d)}{||\mathbf{f}_\text{sam}(\bar{\tau}_d)||_2^2}.
	\end{equation}	
	Based on this, the square error $E_\tau(d,\iota)$ is defined as
	\begin{equation}
		E_\tau(d,\iota)=
		||\hat{\mathbf{x}}_\text{sam}^{(\iota)} - \bar{\beta}_d^{(\iota)} \mathbf{f}_\text{sam}(\bar{\tau}_g)||_2^2.
	\end{equation}
	We find the grid point index that minimizes $E_\tau(d,\iota)$, i.e., $d_\text{opt}^{(\iota)} = \arg\min_{d} E_\tau(d,\iota)$. 
	Therefore, the delay and complex gain estimations of the $\iota$-th path are given by $\hat{\tau}^{(\iota)} = \bar{\tau}_{d_\text{opt}}^{(\iota)}$ and $\hat{\beta}^{(\iota)} = \bar{\beta}_{d_\text{opt}}^{(\iota)}$.
	Finally, we prune estimated scatterers if their complex gains are less than $10\%$ of the maximum complex gain's power.
	This threshold is adopted to prune spurious components with negligible magnitudes (typically observed around $10^{-3}$ to $10^{-4}$ in simulations).
	With the estimated coordinates, delays, and complex gains, the full-dimensional channel can be reconstructed based on \eqref{H}.
	
%

	\begin{figure*}[t]
		\centering
		\subfigure[]{\includegraphics[width=0.46\textwidth]{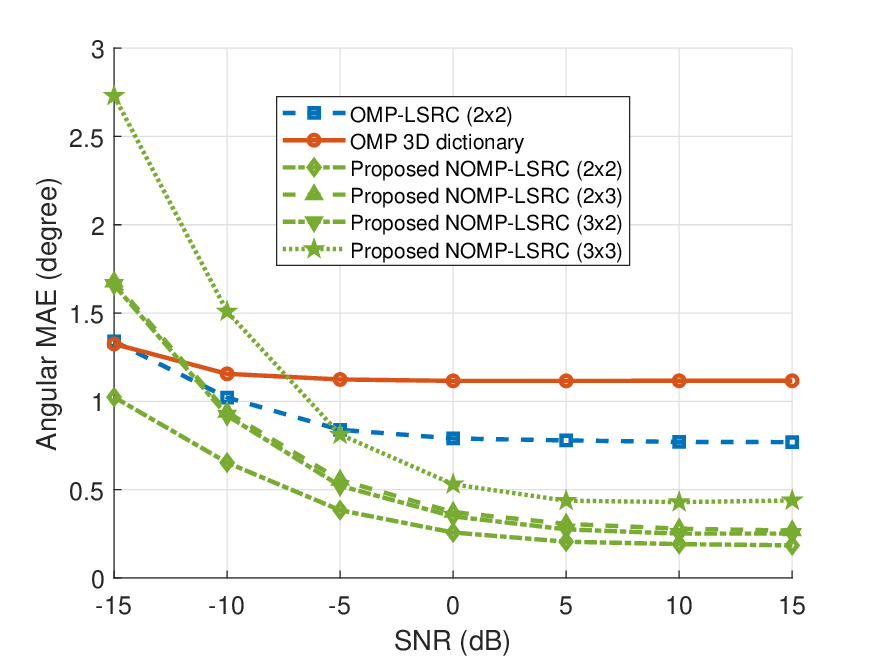}%
			\label{fig:sim2}}%
		\hfill
		\subfigure[]{\includegraphics[width=0.46\textwidth]{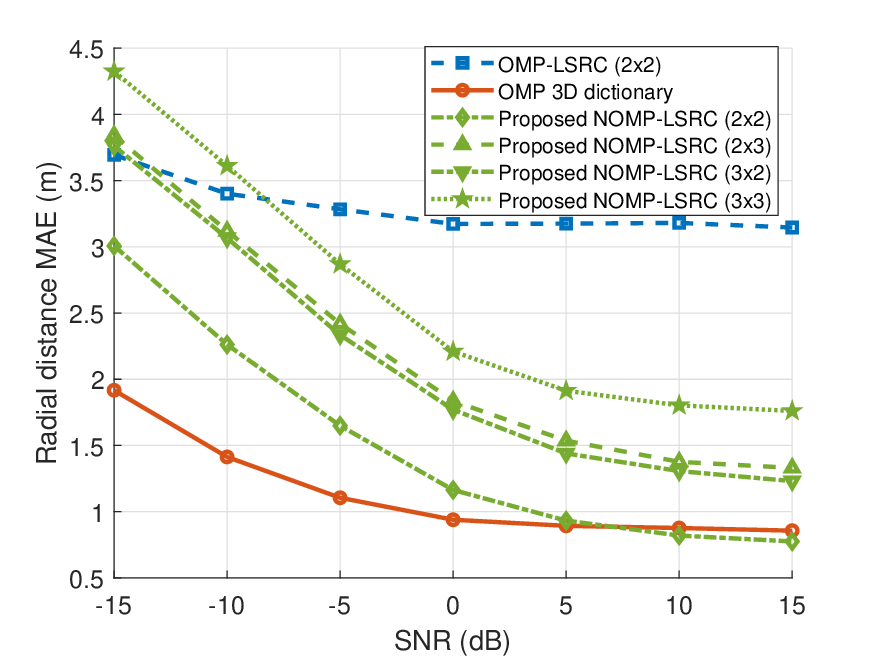}%
			\label{fig:sim3}}%
		\caption{Sensing performance comparison versus SNR for different methods. (a) Angle MAE. (b) Radial distance MAE.}
		\label{fig:sim_set1}
	\end{figure*}
	
	\begin{figure}[t]  
		\centering
		\includegraphics[width=0.46\textwidth]{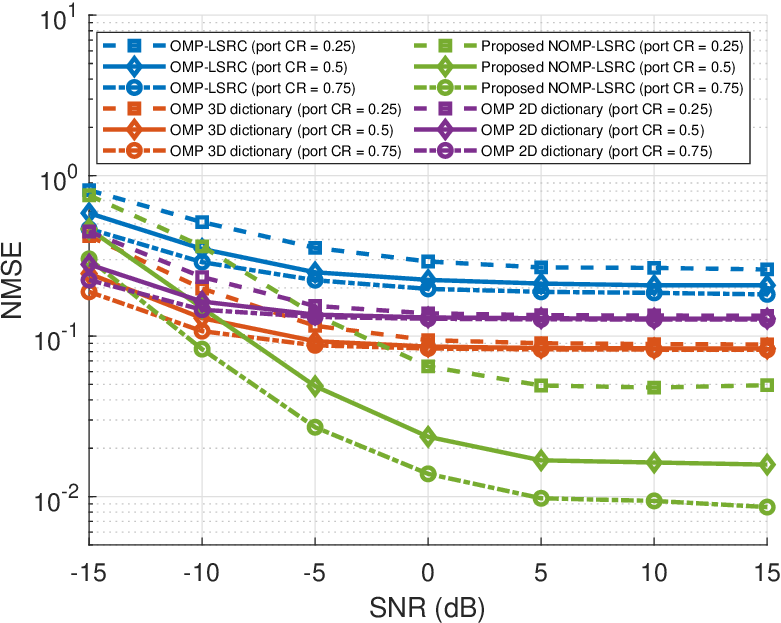}
		\caption{NMSE versus SNR for wideband communication CE.}
		\label{fig:nmse}
	\end{figure}
	
%
%

	\section{Simulation Results}
	
	In this section, we evaluate the communication and sensing performance of the proposed method.
	In the simulation, we set $f_{\text{c}} = 10$\,GHz ($\lambda = 0.03$\,m), $K = 64$, and the subcarrier spacing is 200 kHz.
	The ports for MA are of an $N = 32 \times 32$ uniform planar array shape, with an inter-port spacing of $\frac{\lambda}{2}$. This results in a Rayleigh distance of 15.36 m.
	For the channel model, we set $L = 6$, and the radial distance, azimuth angle, and elevation angle of each scatterer and user are drawn from $\mathcal{U} (2\,\text{m},12.28\,\text{m})$, $\mathcal{U} (\frac{\pi}{6},\frac{5\pi}{6})$, $\mathcal{U} (\frac{\pi}{6},\frac{5\pi}{6})$, respectively.
	Delays are decided based on the location of user and scatterers.
	For the subregion division, the whole region is partitioned into a grid of $Q=N_x \times N_z$ subregions, where $N_x$ and $N_z$ represent the number of uniform divisions along the $x$-axis and $z$-axis, respectively. 
	For the proposed scheme, we randomly activate $N_\text{T}$ ports for each subregion in one frame, and all subregions share the same pilot pattern. In this situation, $K_{\text{c}} = 32$, $L_{\text{pre}}=10$, $G_\theta=60$, $G_\phi=30$, $R=10$, $\alpha_\text{th}=10^\circ$, $\tau_\text{max} = 115$ ns, and $G_\tau=400$. We set $Q=2\times 2$, and $N_{\text{T}} = 128$ unless stated otherwise.
	The mean absolute error (MAE) of angle and distance and normalized mean squared error (NMSE) of channels are adopted for sensing and CE evaluation, respectively.
	
	In Figs. \ref{fig:sim_set1} and \ref{fig:nmse}, we show the NMSE and sensing metrics versus signal-to-noise ratio (SNR).
	Specifically, we consider 4 different CE or sensing methods for comparison: {\bf 1) OMP-LSRC}: A subregion-based method utilizing OMP 2D dictionary to estimate elevation and azimuth angles and LSRC for localization. 
	{\bf 2) OMP 3D dictionary}: A full-region-based method utilizing OMP 3D dictionary to estimate 3 position parameters of scatterer.
	{\bf 3) Proposed NOMP-LSRC}: A subregion-based method utilizing NOMP to estimate elevation and azimuth angles and LSRC for localization. 
	{\bf 4) OMP 2D dictionary}: A full-region-based method utilizing OMP 2D dictionary to estimate 2 angle parameters of scatterer, discarding radial distance.
	Fig. \ref{fig:nmse} shows that our proposed method can achieve lower NMSE at high SNR.
	This is because we obtain more accurate angular estimate from NOMP and the precise location of scatterer estimate from LSRC. 
	Besides, it is shown that NMSE performs better if a higher MA ports compressed ratio (CR), defined by $M/N$, is considered.
	For sensing performance, Figs. \ref{fig:sim2} and \ref{fig:sim3} compare angle MAE and radial distance MAE, respectively. 
	Observe that our method shows a significant performance improvement in terms of all considered metrics.
	In addition, we simulate different subregion divisions with different $Q$. 
	Figs. \ref{fig:sim2} and \ref{fig:sim3} show that our proposed method performs best for a $2\times 2$ subregion division.
	
\begin{figure}[t]
		\centering
		
		\subfigure[]{\includegraphics[width=0.48\linewidth]{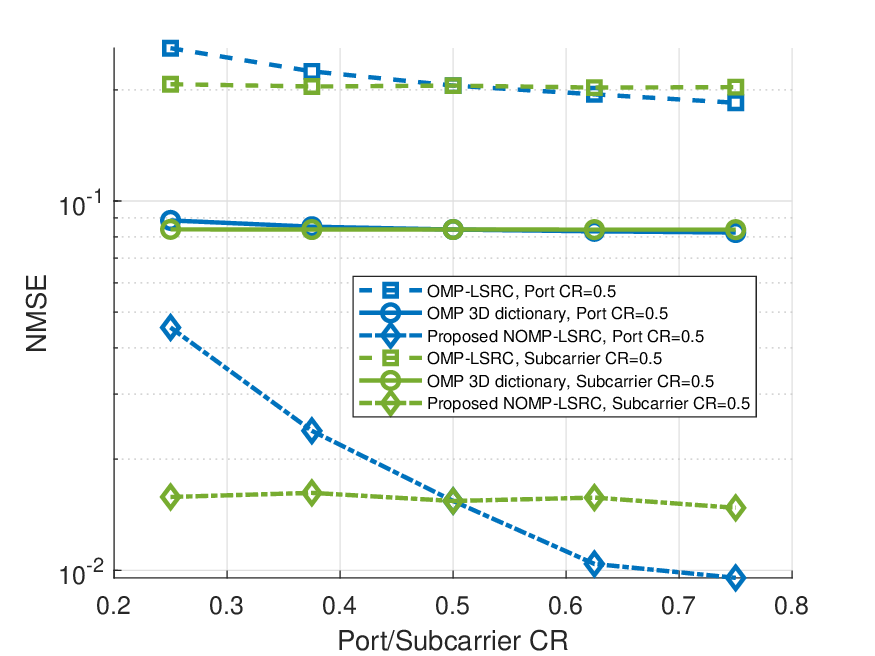}%
			\label{sim_ce}}%
		\hfill 
		\subfigure[]{\includegraphics[width=0.48\linewidth]{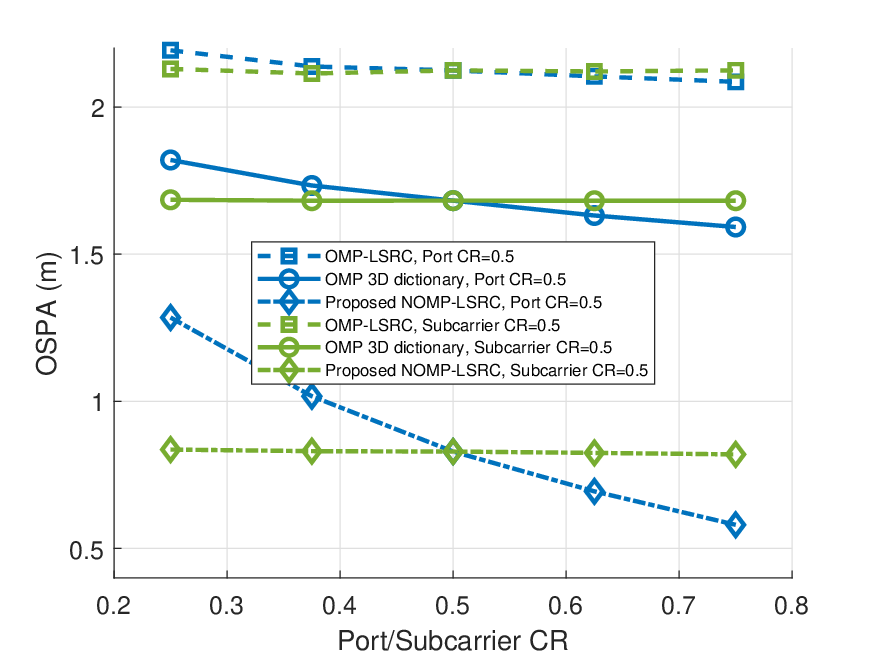}%
			\label{sim_se}}
		
		\caption{CE and sensing performance comparison versus CR for different methods. (a) NMSE. (b) OSPA.}
		\label{fig:sim_cr}
	\end{figure}

	To evaluate the system performance under different resource constraints, we investigate the impact of CRs in both ports and subcarriers. We define  the subcarriers CR as $K_\text{c}/K$, representing the pilot overhead. Fig. \ref{fig:sim_cr} illustrates NMSE and the optimal subpattern assignment (OSPA) against the ports and subcarriers CR with SNR $=20$ dB. 
	Note that, the definition of OSPA is given in Appendix \ref{appe}, taking into account both false alarm or missed detection (i.e., $N_\text{clu} \neq L$).
	It is observed that increasing the ports measurements significantly enhances the CE/sensing accuracy due to the higher spatial resolution provided by more MA ports. 
	In contrast, it is shown that the system performance is insensitive to the subcarriers CR. Even with a low ratio, the proposed scheme maintains high accuracy.
	This is because the proposed NOMP-based angle estimation exploits the MMV structure, where the angular support of scatterers is common across all subcarriers. This shared sparsity allows the algorithm to effectively identify scatterer directions even with limited frequency domain samples.
	
	\section{conclusion}
	
	In this paper, we proposed an ISAC framework for near-field MA-aided communication systems, which can reconstruct complete CSI and locate scatterers effectively.
	To achieve the estimation of the scatterer coordinates, we divide the MA movement area into several subregions to estimate angles separately, and then utilize geometric relationships to locate scatterers jointly.
	Through this method, we refined MA CE utilizing the estimated scatterer locations.
	The simulation results demonstrated that our proposed method leads to improvement over the traditional CS method in terms of both scatterer localization and communication channel acquisition.

	\begin{appendices}
		\section{OSPA Metric}
		\label{appe}
		The OSPA metric measures the difference between a true set of targets, $\mathcal{X}= \{\mathbf{x}_1, \dots, \mathbf{x}_m\}$, and an estimated set, $\mathcal{Y}= \{\mathbf{y}_1, \dots, \mathbf{y}_n\}$. Assuming $m \le n$ (otherwise the sets are swapped), the OSPA distance is defined as\cite{OSPA}
		\begin{equation*}
			\begin{aligned}
					D(\mathcal{X},\mathcal{Y}) = \frac{1}{n} \left(
					\min_{\pi \in \Pi_n} \sum_{i=1}^{m} 
					d_\psi \left(\mathbf{x}_i, \mathbf{y}_{\pi(i)}\right) + \psi(n-m)
					\right),
			\end{aligned}
		\end{equation*}
		where $\psi > 0$ is the cutoff parameter defining the maximum localization penalty. 
		We set the cutoff parameter for successfully clustered scatterers is $\psi=3$ (meter).
		The term $d_\psi(\mathbf{x}, \mathbf{y}) = \min(\psi, \|\mathbf{x}-\mathbf{y}\|_2)$ is the Euclidean distance truncated at $\psi$. $\Pi_n$ is the set of all permutations of $\{1, \dots, n\}$, and $\pi$ is a specific permutation used to match targets. The first part of the formula calculates the localization error for the optimal assignment, while the second part penalizes the cardinality error (unmatched points).
	\end{appendices}


	\bibliographystyle{IEEEtran}
	\bibliography{references}
	
\end{document}